\documentclass[superscriptaddress,showpacs,twocolumn,amssymb,aps,pre,10pt]{revtex4-1}

\usepackage{graphicx,dcolumn,bm,amsmath,color}

\newcommand{\equ}[1]{Eq. ~(\ref{#1})}

\newcommand{\eeq}{ \end{equation} }
\newcommand{\beq}{ \begin{equation} }
\newcommand{\bea}{\begin{eqnarray}}
\newcommand{\eea}{\end{eqnarray}}

\newcommand{\ga}{\alpha}

\newcommand{\gl}{ \lambda }

\newcommand{\bhu}{ \hat{\bf u} }

\newcommand{\bbr}{ {\bf r} }

\begin{document}

\title{Capturing self-propelled particles in a moving microwedge}

\author{A. Kaiser}
\email{kaiser@thphy.uni-duesseldorf.de}
\affiliation{Institut f\"ur Theoretische Physik II: Weiche Materie,
Heinrich-Heine-Universit\"at D\"{u}sseldorf, 
Universit{\"a}tsstra{\ss}e 1, D-40225 D\"{u}sseldorf, Germany}
\author{K. Popowa}
\affiliation{Institut f\"ur Theoretische Physik II: Weiche Materie,
Heinrich-Heine-Universit\"at D\"{u}sseldorf, 
Universit{\"a}tsstra{\ss}e 1, D-40225 D\"{u}sseldorf,
Germany}
\author{H. H. Wensink}
\affiliation{Laboratoire de Physique des Solides, Universit\'{e} Paris-Sud and CNRS, B\^{a}timent 510, 91405 Orsay Cedex, France}
\author{H. L\"{o}wen}
\affiliation{Institut f\"ur Theoretische Physik II: Weiche Materie,
Heinrich-Heine-Universit\"at D\"{u}sseldorf, 
Universit{\"a}tsstra{\ss}e 1, D-40225 D\"{u}sseldorf,
Germany}

\date{\today}

\begin{abstract}
Catching fish with a fishing net is typically done either by dragging a fishing net through quiescent water or by placing a stationary
basket trap into a stream. We transfer these general concepts to micron-sized self-motile particles moving in
a solvent at low Reynolds number and study their collective trapping behaviour  by means of  computer simulations of a two-dimensional system of self-propelled rods. A chevron-shaped obstacle is dragged through the active suspension with a constant speed $v$ and acts as a trapping ``net". Three trapping states can be identified corresponding to no trapping, partial trapping and complete trapping and their relative stability
is studied as a function of the apex angle of the wedge, the swimmer density and the drag speed $v$. When the net is dragged along the inner wedge, complete trapping is facilitated and a partially trapped state 
changes into a complete trapping state if the drag speed exceeds a certain value.  Reversing the drag direction leads to a reentrant transition from no trapping,  complete trapping, back to no trapping upon increasing the drag speed along the outer wedge contour. The transition to complete trapping is marked by a templated self-assembly of rods forming polar smectic structures  anchored onto the inner contour of the wedge. Our predictions can be verified in experiments 
of artificial or microbial swimmers confined in microfluidic trapping devices.
\end{abstract}

\maketitle
\section{Introduction}
\label{sec:intro}
With an appropriate use of a fishing net, many fish can be simultaneously
caught in an efficient way. There are two different strategies to catch
fish using e.g. a cone-shaped net; the net can either be dragged through quiescent water or a
stationary trap (a so-called "fyke") can be placed in running water forcing the fish to swim inside the fyke. While the general methods for trapping macroscopic swimming organisms (fish) have been known since ancient times
\cite{fishing}, the corresponding problem in the microscale has  been scarcely explored thus far owing to general difficulty in controlling and designing processes in systems of micron-sized objects.  There
are many realizations of microscopic swimmers  \cite{Cates_2012,Romanczuk2012,Marchetti_Rev,Vicsek_Report2012}
including autonomously navigating microbes \cite{2004DoEtAl,2007SoEtAl,2012SwinEtAl,GoldsteinPRL2012,Poon,DiLuzio2005,DiLuzio2006,Hill_PRL2007,Shenoy_PNAS,Schmidt2008,Garcia}
and man-made artificial swimmers  
\cite{Golestanian_PRL05,Baraban2008,Bocquet_PRL10,Bibette,Ignacio,Kaehr,Bechinger_SM11,snezhko_prl,snezhko_nature,2007Kapral,Ohta,BtH_L_part,Reinmueller,Living}.  For many
applications it is of key importance to trap collections of these active
particles into a moving trap. A first application is to transport
ensembles of swimmers to a given destination like a cargo. This situation differes from the more commonly considered case in which the swimmer itself transports an inert cargo
\cite{Raz,Gao,Baraban_cargo,Dietrich_EPL,AngelaniCARGO}. It is obvious that, in the former situation, one first has to catch the particles in an
efficient and controlled way before they can be transferred to the specific destination via a moving trap. A second application could be 
to efficiently remove "dangerous" toxic particles in order to clean
the environment \cite{contamination1,contamination2}. Also here, the motion of the trap is expected to play a crucial role in optimizing the removal of contaminating mesogens.

Apart from the different length scale, 
it is important to note that another basic difference between macroscopic
fish and microbes is their Reynolds number $\text{Re}$ which characterizes the
ratio of inertial to viscous forces accociated with swimming. While fish typically
swim at Reynolds number of a several hundred, micro-swimmer typically operate at very low Reynolds numbers, $\text{Re} \ll 1$.

In this paper, we  transfer the ideas of catching fish in a net to
micron-sized self-motile particles propagating through a solvent at
low Reynolds number. We use computer simulations of a two-dimensional
system of self-propelled rods and drag a chevron-shaped obstacle with a
constant speed $v$ through the embedding active fluid.  As revealed by a simple
Galilean transformation, this set-up is equivalent to a static trap in a
flowing solvent.
Our simulations complement earlier studies
for a static trap \cite{Kaiser_PRL} where a wedge was found to
optimize the catching efficiency. Here, we focus on the effect of a nonzero
drag speed. At fixed
swimmer density and varied drag velocity $v$ and apex angle of the trap,
there are three emerging states corresponding to no trapping,
 partial trapping and complete trapping. While in the no trapping
state, no particles remain in the trap over time, in the
complete trapping state all swimmers
are permanently caught in the microwedge after a certain amount of  time. Finally
 partial trapping is referred to a state where only a
fraction of the particles is permanently trapped. Obviously, the
dream of any fishermen and the most desirable situation in many applications
is the complete trapping state where no freely
moving particles are left.

We solve the single rod case analytically and present the trapping state diagram in the plane
spanned by the opening angle $\alpha$ of the microwedge ($0<\alpha\leq \pi$) and the trap velocity $v$
(normalized by the swimmer velocity $v_0$). The drag direction is along the symmetry axis of the
wedge and we define a positive drag speed if its drag is along the inner part of the wedge.
 As a result, if the net is dragged into the positive
direction, trapping is facilitated. Counterintuitively, however, for a negative drag velocity,
a situation of no trapping can change into a trapped one which we attribute to polar ordering
of the swimmer along the wedge symmetry axis.  Clearly, when the (negative) trap velocity exceeds
the swimmer velocity ($v/v_0 <-1$), trapping is not any longer possible
as the trap overtakes the swimmer which leads to a reentrant effect for increasing negative velocity:
 for intermediate opening angles $\alpha$, we observe the state sequence no trapping, complete trapping,  no trapping.
For finite trap density we employ computer simulations \cite{Kaiser_PRL,Wensink2008,Borge_PRE_2011}
and confirm the trends of single-particle trapping. For high enough positive drag speed, a partial trapped situation will
change into a complete trapped situation. In the converse case of a negative trap velocity, the reentrance effect
is amplified by a collective polar ordering
in the trap. This is a typical example of self-assembly of self-propelled colloidal rods \cite{FurstRev} 
directed by the moving microwedge.
Previous studies analysing the rectification effect
of a wall of funnels by experiments \cite{Chaikin,DunkelPNAS}, theory \cite{CatesTrapping} and simulation \cite{Reichhardt_PRL2008,Reichhardt,WanSoftMatter2013,Ghosh} have utilized similar chevron-shaped boundaries
 but never focussed on trapping.

Apart from their relevance for applications, our predictions can be
verified in experiments on rod-like microbes and self-propelled colloids and granulates \cite{Chaikin,LeonardoPNAS,SokolovPNAS,Aranson_RMP,Ramaswamy}. Typically
the system is moving on a two-dimensional substrate or can be subject to a
strong two-dimensional confinement \cite{Wensink_PNAS}. A chevron-like trap can be
prepared by lithographic techniques \cite{Kaehr,Bechinger_SM11,
microfluidic,Mino_Clement} and it can be dragged at wish
using optical tweezers \cite{Aranson_SM12,Bechinger}. Therefore an experimental
realization of our model is conceivable. We further anticipate that the
same effects occur also in three dimensions where the corresponding
generalization of the wedge-like trap is a hollow cone - similar in spirit
to a real fishing net.

This paper is organized as follows: in Section \ref{sec:model} we introduce the model and
explain the simulation method.  Section \ref{sec:single} is devoted to the case of
a single self-propelled rod. We will give a theoretical prediction of the trapping state diagram along with numerical results. 
In Section \ref{sec:many}, we investigate the trapping states for many particles
and all three main control parameters. In particular, we fix each time one of these, vary the
others and obtain a full trapping state diagram which can be explained by the effects already showing up for the 
single particle case. 
Finally, we conclude in Section \ref{sec:conc}.

\section{Model}
\label{sec:model}
The aim is to formulate a minimal collision-based model for self-propagating rod-shaped particles that is capable of capturing the generic features of interacting swimmers at intermediate to high particle density and their collective response to mobile confining boundaries.
Following earlier studies \cite{Kaiser_PRL,WensinkJPCM}, our system consists of 
$N$ rigid rods of length $\ell$, 
each moving in the overdamped limit with a propagation velocity $v_0$ arising from a formal  
force $F_0$ fixed along the longitudinal rod axis $\bhu$
\cite{WensinkJPCM,Wensink2008}.
This does not contradict the basic fact that a swimmer is force-free.
The actual position of the $\alpha$th rod $(\ga=1,\ldots,N)$ is described
by a centre-of-mass position vector $\bbr_{\ga}$ and a unit orientational
vector $\bhu_{\ga} = (\cos \varphi_{\ga}, \sin \varphi_{\ga})$.

Due to solvent friction the particles move in the overdamped
low Reynolds number regime, while interacting with the other particles and the
boundary by steric forces only \cite{Wensink2008}. The latter are implemented by
discretizing each rod into a linear array of $n$ equidistant spherical segments and imposing a repulsive
Yukawa potential  between the
segments of each pair \cite{Kirchhoff1996,Graf_Loewen_PRE_1999}. The total pair potential
between rods $\{\alpha, \beta \}$ with orientational unit
vectors $\{ \bhu _{\alpha}, \bhu _{\beta} \}$ and centre-of-mass
distance ${\Delta \bf r}_{\alpha \beta} = \bf r_{\alpha} - \bf r_{\beta} $ is then given by 
\bea
U_{\alpha \beta} = U_{0} \sum_{i=1}^{n}\sum_{j=1}^{n}
\frac{\exp [-r_{ij}^{\alpha \beta} / \lambda] }{ r_{ij}^{\alpha \beta}}
\eea
where $U_{0}>0$ defines the amplitude, $\lambda$ the screening length and
$r_{ij}^{\alpha \beta} = |{\Delta \bf  r}_{\alpha \beta} + (l_{i}
\bhu_{\alpha} - l_{j} \bhu_{\beta})|$ the distance
between segment $i$ of rod $\alpha$ and $j$ of rod $\beta$ ($\alpha
\neq \beta $) with $l_{i} = d(i-1)$, $i \in [1,n]$
denoting the segment position along the main rod axis. The number of rod segments $n$ is 
chosen such that the intrarod segment distance $d = \ell / ((n+1)(n-1))^{1/2} \leq \gl$ 
and rod overlaps are prevented. A trap is introduced as a boundary with a
prescribed shape and contour length $\ell_{T}$ and is dragged with a 
velocity $v$ through the system. Particle-trap interactions are
implemented by discretizing the trap boundary into $n_{T} = \lfloor \ell_{T}/d \rceil $ equidistant
segments each interacting with the rod segments via the same Yukawa potential, resulting in the pair potential
\bea
U_{\alpha T} = U_{0} \sum_{i=1}^{n}\sum_{k=1}^{n_{T}}
\frac{\exp [-r_{ik}^{\alpha T} / \lambda] }{ r_{ik}^{\alpha T}}.
\eea
Here $r_{ik}^{\alpha T}$ denotes the distance between segment $i$ of rod $\alpha$ and segment $k$ of the
trap.
The net is dragged with imposed velocity $\textbf{v} = v \textbf{x}$ along the symmetry axis of the wedge 
according to
\bea
\textbf{r}_{k} = \textbf{v}t,
\eea
where $\textbf{r}_{k}$ denotes the position of the $k$th segment of the trap.
Mutual self-propelled rod collisions generate apolar nematic alignment which stimulates swarm formation at finite 
concentrations \cite{Ramaswamy}. The wedge boundary mimics a hard wall and imparts 2D planar 
order with rods pointing favorably perpendicular to the local wall normal.

The microscopic equations of motion for the centre-of-mass position $\bbr_\ga(t)$ and orientation
$\bhu_{\ga}(t) = (\cos \varphi_{\ga}(t),\sin \varphi_{\ga}(t))$ of the self-propelled particles emerge
from a balance of the forces and torques acting on each rod $\ga$ and are
similar as described in Ref.~\cite{WensinkJPCM}
\bea
\label{e:eom_r}
{\bf f }_{\mathcal{T}} \cdot \partial_{t}\bbr_{\ga}
&=& -\nabla_{{\bbr}_{\ga}} U + F_0 \bhu_{\ga} ,
\\
{\bf f}_{\mathcal{R}} \cdot \partial_{t} \bhu_{\ga}
&=&
-\nabla_{\bhu_{\ga}} U,
\label{e:eom_u}
\eea
in terms of the total potential energy  $U=(1/2)\sum_{\alpha, \beta (\alpha \neq
  \beta)} U_{\alpha \beta} + \sum_{\alpha ,T} U_{\alpha T}$
with $U_{\alpha T}$ the potential energy of rod $\alpha$ with the
trap and $\nabla_{\bhu_{\ga}}$ denotes the gradient on a unit circle. The one-body translational and rotational 
friction tensors ${\bf f}_{\mathcal{T}} $ and ${\bf f}_{\mathcal{R}}$ can be decomposed into parallel 
$f_{\parallel}$, perpendicular $f_{\perp}$ and rotational $f_{\mathcal{R}}$ contributions
\bea
{\bf f}_{\mathcal{T}} &=&
f_0 \,\left[ f_{\parallel} \bhu_{\ga} \bhu_{\ga} + f_{\perp} ({\bf I}
- \bhu_{\ga} \bhu_{\ga}) \right],\\
{\bf f}_{\mathcal{R}} &=& f_0\, f_{\mathcal{R}} {\bf I}.
\eea
Hereby $\bf I$ is the 2D unit tensor and $f_0$ is a Stokesian friction coefficient. 
The dimensionless geometric factors
$\{ f_{\parallel}, f_{\perp}, f_{\mathcal{R}} \}$ depend solely on the
aspect ratio $a=\ell/\lambda$, and we adopt the standard expressions for rod-like macromolecules,
as given in Ref.~\cite{tirado}
\bea
f_{||} = 2\pi \left(\ln a - 0.207 + 0.980a^{-1} - 0.133a^{-2}\right)^{-1},
\nonumber \\
f_{\perp}= 4\pi \left(\ln a+0.839 + 0.185a^{-1} + 0.233a^{-2}\right)^{-1},
\nonumber \\
f_{\mathcal{R}} = \frac{\pi a^2}{3}\left( \ln a - 0.662 + 0.917a^{-1} - 0.050a^{-2}\right)^{-1}.
\eea
\equ{e:eom_u} neglects thermal or intrinsic Brownian noise \cite{DrescherPNAS}, which is
acceptable at intermediate-to-high concentrations when
particle collision dominate the dynamics \cite{Wensink_PNAS}. Despite its minimal nature the self-propelled rod model provides a remarkably accurate description of the velocity statistics and microstructure of dense active matter \cite{Wensink_PNAS}.

It is important to note that our system is strictly equivalent to a quiescent net where the
swimmers all experience their propagation velocity $\textbf{v}_{0}$ together with an added velocity $-\textbf{v}$.
This can easily be demonstated by transforming the equation of translational  motion \equ{e:eom_r} in terms of reduced difference 
coordinates $\widetilde{\textbf{r}}_{\alpha}=\textbf{r}_{\alpha}-\textbf{v}t$ i.e. by applying
a \emph{Galilean transformation} \cite{Lowen_Hoffmann} so that:
\bea
\partial_{t} \widetilde{\textbf{r}}_{\ga} = ({v}_0\bhu_\ga - \textbf{v}) - {\bf f }_{\mathcal{T}}^{-1} \cdot \nabla_{{\widetilde{\textbf{r}}}_{\ga}} U.
\eea
The typical self-propulsion speed of a free single self-propelled rod 
\bea 
v_{0}=\frac{F_0}{{f_{0}f_{||}}}
\eea 
defines the time interval  
\bea
\tau = \frac{\ell}{v_{0}} 
\eea
a rod needs to swim a distance comparable to its size.
In the following, distances are measured in units of $\ell$ and energy in units of $F_0\ell$.

\begin{figure}[t]
\centering
\includegraphics[clip=,width= 0.72 \columnwidth]{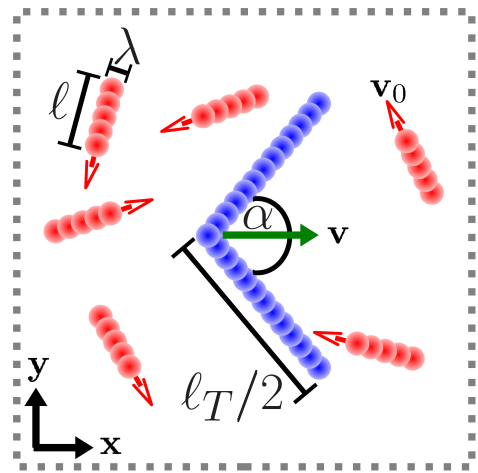}
\caption{(color online) Sketch of the system of self-propelled rods with aspect ratio $a = \ell / \lambda$ and an self-motile velocity \textbf{v$_{0}$} directed along the main axis $\bhu$ (red (light gray) arrows) of each rod at bulk density $\phi$. The single Yukawa segments are shown as red spheres. A mobile wedge (indicated by blue (dark) spheres) with contour length $\ell_{T}$ and an apex angle $\alpha$ is dragged with a constant velocity \textbf{v} (green (filled gray) arrow). Periodic
boundary conditions in both Cartesian directions are indicated by dotted lines.
\label{f1}
}
\end{figure}

We simulate self-propelled rods with aspect ratio $a=10$, using $n=11$ segments, in a square simulation box with area
$A$ and periodic boundary conditions in both Cartesian
directions.  A particle packing fraction is defined as $\phi = N\sigma/A$ with
$\sigma= \lambda(\ell-\lambda) + \lambda^{2}\pi/4$ the effective area of
a single rod. In the bulk density range $\phi <0.2$ the self-propelled rods spontaneously form flocks
with strong spatial density fluctuations \cite{Narayan}.  
We subject the self-propelled rods to a mobile chevron boundary with
contour length $\ell_{T} = 20 \ell$ and variable apex angle $0^{\circ} <
\alpha < 180^{\circ}$, leading to an occupied trap area $A_0 = (\ell^2_T /8) \sin \alpha/2$, 
which is dragged with velocity $v$ (see Fig.~\ref{f1}). In the macroscopic limit, 
the system can be interpreted  as a reservoir of microswimmers exposed to an equidistant array of
mutually independent wedges. A reduced trap density is defined by
$\phi_{T} = (\ell^2_T /8A)$ which fixes the number of rods via 
$N=(\ell_{T}^{2}/8 \sigma)(\phi/\phi_{T})$. We constrain $\phi_{T} = 0.031 < 0.1$
in order to guarantee the microwedges to be completely independent of each other within
the typical range of bulk rod packing fractions  $0 < \phi < 0.1$ considered here.
The latter density is one of our main steering parameters. There are also the 
drag velocity which is in the range of $-1.2v_{0} < v < 8v_{0}$  and
the apex angle $\alpha$ of the microwedge.
 
Initial configurations are generated from a rectangular lattice of aligned
rods with $\bhu$ pointing randomly up or down. 
The rods are randomly displaced from the initial lattice such that the starting
configuration bears already some randomness. The segments of the microwedge are successively
placed in the system to avoid overlapping rods. 
We simulate the whole system for a time of at least $t_{s} = 15000 \tau$.

\section{Trapping a single swimmer in a mobile microwedge}
\label{sec:single}

\begin{figure*}[t]
\centering
\includegraphics[clip=,width= 1.5 \columnwidth]{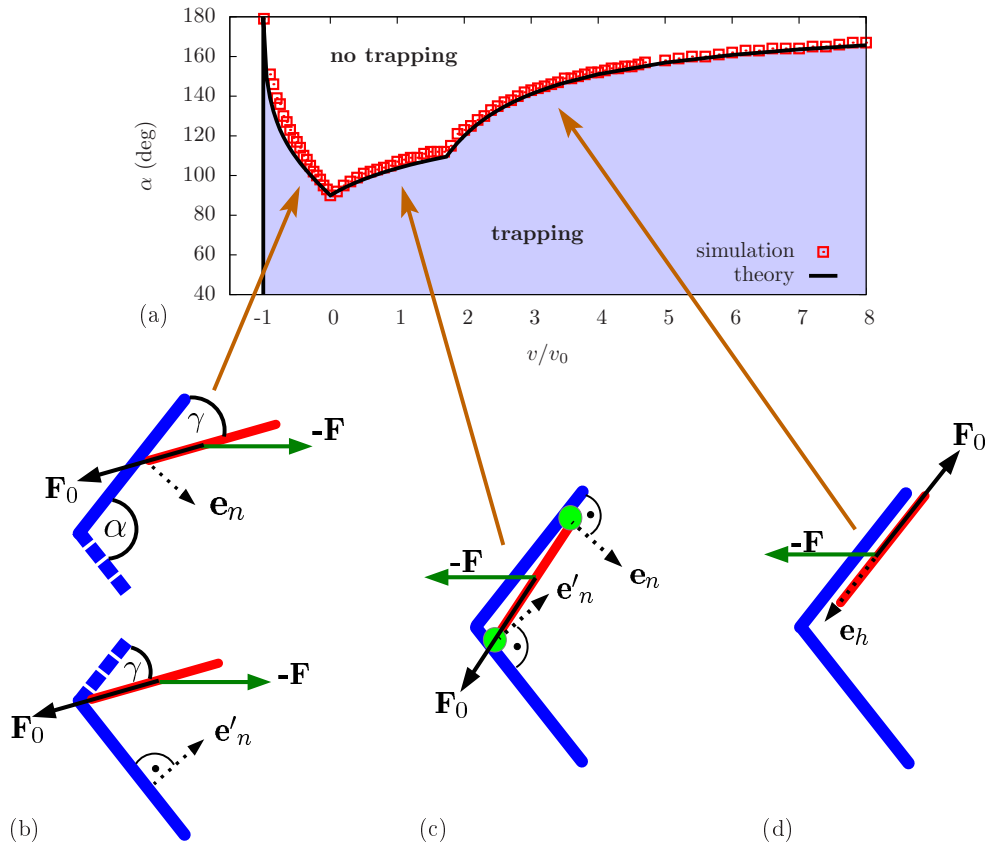}
\caption{(color online) (a) Trapping state diagram for a single self-propelled rod in the plane of
reduced drag velocities $v/v_{0}$ and net apex angles $\alpha$. The shaded region marks the trapping regime. The dots represent simulation
results for the trapping- no trapping-boundary while the solid line contains the analytical predictions. Different
 trapping mechanisms are sketched in (b)-(d). For more details, see text. Points of contact of the swimmer and the microwedge are highlighted by light green (light gray) circles.
\label{f2}
}
\end{figure*}

We first focus on a single swimmer for which analytical results can be obtained 
which we test against our computer simulations.
 In Fig.~\ref{f2}(a), simulation results and analytical formulae for the trapping state diagram are combined.
The main control parameters we vary are the reduced trap drag velocity $v/ v_{0}$ and the apex angle $\alpha$. 
The trapping scenario of a single swimmer is generic and is independent of the contour length of the net, as long as $\ell_{T} \gg \ell$, and the aspect ratio of the rod-shaped swimmer.
In the simulation, a particle is considered to be trapped if it remains inside the wedge for at least $t^{\ast} = 10^{3} \tau$. 

Let us first discuss some limiting cases which are all intuitive: For strongly negative drag velocities, $v/v_{0}<-1$, 
the swimmer is slower than the microwedge and can therefore never get trapped for any opening angle $\alpha$.
Conversely, for $v/v_{0}>-1$ and very small opening angles, once a rod enters the moving net it is faster than the net
and will therefore approach to the kink of the wedge where it stays for ever since it cannot escape by turning around.
Hence there is a trapping state
 for $v/v_{0}>-1$ and small opening angles. Complementarily, for $v/v_{0}>-1$ and very large opening angles
($\alpha \approx \pi$), if the rod enters the microwedge, it will just slide along the wall of the wedge and will then
pass over the slight kink of the wedge such that the leaves the trap again. Consequently, the rod does not permanently reside in the wedge and thus attains a no trapping state. 

As shown in Fig.~\ref{f2}(a), the intermediate transition opening angle 
which separates the trapping from the no trapping regime is a marked function of the reduced trap velocity
which exhibits some cusps. The cusps occur at $v/v_{0}=-1$, $v/v_{0}=0$, and $v/v_{0}=\sqrt{3}$ 
and clearly distinguish four different situations which we now discuss quantitatively step by step.
We use the frame of the resting net for this discussion and introduce forces instead of velocities.
Clearly forces are proportional to velocities. In the microwedge system, the rod center experiences a force 
${\bf F_{0}} \propto {\bf v}_{0}$ governing its self-propulsion 
plus another force $-{\bf F} \propto -{\bf v}$ arising from the resting rod frame. Third the wall possibly exerts
at contact a force ${\bf F}_{N}$ onto the rod which is always normal to the wall.

As already stated above, for strongly negative  drag velocities, $v/v_{0}<-1$, 
a swimmer moves slower than the microwedge and can therefore never get trapped. For $-1<v/v_{0}<0$, 
it is expected that single rods are still spilled out by the net, but the opposite behaviour is true: trapping is getting more efficient
if the drag speed approaches the swimmer speed $v/v_{0}\to -1^{+}$ from above. This counter-intuitive behaviour can be 
understood as sketched in Fig.~\ref{f2}(b). If a rod enters the trap and hits a wall (see upper sketch of Fig.~\ref{f2}(b) and \cite{SuppMat}), 
it will orient at an angle $\gamma$. This angle is determined by the condition that 
 the projection of ${\bf F_{0}}-{\bf F}$ onto the wall normal
has to vanish, $(\textbf{F}_0-\textbf{F}) \cdot \textbf{e}_n =0$, with the wall normal vector 
$\textbf{e}_n = (\sin \alpha/2, -\cos \alpha /2)$. This leads to $\gamma = \arcsin (v/v_0)$. With this orientation, the rod will 
 slide along the wall inside the corner until it touches the lower wedge wall [see lower sketch of Fig.~\ref{f2}(b)]. The threshold
condition whether the rod slides further outside the wedge along the lower wall is finally given by requiring that the normal projection 
along the lower wall normal $\textbf{e}'_n = (\sin \alpha/2, \cos \alpha /2)$ vanishes, i.e.:
 $(\textbf{F}_0-\textbf{F}) \cdot \textbf{e}'_n =0$. This alltogether leads to the threshold condition
\bea
\alpha =  \frac{\pi}{2} - 2 \left( \arcsin \left( \frac{v}{\sqrt{2}v_{0}} \right) - \arcsin \left( \frac{v}{v_{0}} \right)   \right).
\eea 
which is plotted as a solid line in Fig.~\ref{f2}(a). This implies that a trap moving into the negative direction orients the rods along the wedge symmetry axis
and thus keeps them inside, enhancing thereby the trapping efficiency.

In case of positive drag velocities, two different trapping mechanisms can occur. The first mechanism is shown in Fig.~\ref{f2}(c) and \cite{SuppMat}.
Here the swimmer enters into the wedge and is stuck close to the wedge cusp with two contact points, one at the highter and another at the lower wall.
This hinders the rod in rotating further such that it gets immobilized.
The contact points are indicated in Fig.~\ref{f2}(c). In this situation, the rod experiences four different forces, two arising 
simultaneously from the wall normals plus  ${\bf F_{0}}-{\bf F}$. The normal wall forces are unknown but  determined by
the joint conditions of vanishing total force and torque acting on the rod center which are characterizing the transition from
 no trapping to trapping.
Hence, these conditions are
$\textbf{F}_0 - \textbf{F} + \textbf{F}_n + \textbf{F}'_n=0$ and 
${F}_n \textbf{e}_h \times \textbf{e}'_n + {F}'_n \textbf{e}_h \times \textbf{e}_n  = 0$. 
Eliminating the unknown normal forces, we obtain the threshold criterion
\bea
\frac{v}{v_{0}} = - \cos \left( \frac{\alpha}{2} \right) + \frac{\sin ^{4} \left(\alpha/2 \right)}{\cos ^{3} \left(\alpha/2 \right)}.
\eea

The second mechanism is shown in Fig.~\ref{f2}(d) (see also \cite{SuppMat}) and refers to a situation where an aligned rod intends to leave the trap, for example when it was able
to turn in the kink. If the projection of $-{\bf F}$ tangential to the wall exceeds the self-propulsion, the moving microwedge surpasses
the rod and keeps it caught. The condition for the threshold for this second mechanism is therefore  $\textbf{F}\cdot \textbf{e}_{h} = F_0$
for   $\textbf{e}_{h} = (-\cos \alpha /2, -\sin \alpha /2)$ which yields
\bea
\cos \left( \frac{\alpha}{2} \right) = \frac{1}{v/v_{0}}.
\eea
As can be shown easily, this second mechanism surpasses the former mechanism for dragging velocities $v/v_{0} > \sqrt{3}$.

Summarizing, a single self-propelled particle can always be trapped for $v > -v_0$ and $\alpha < 90 ^{\circ}$.
For apex angles larger than $90 ^{\circ}$ and increasing trap velocities, 
 the following sequence of states is found: no trapping, trapping, no trapping, trapping.
This clearly demonstrates the nontrivial interplay between wedge geometry and orientational coupling to the rod.
Moreover we find for positive drag velocities two different mechanisms which hold the particles  inside the microwedge.
Finally, the good agreement of the threshold lines between analytical theory and simulations shows that the segment model used in this 
paper reproduces well the purely geometric conditions of almost hard interactions.

\section{Collective trapping}
\label{sec:many}

\subsection{Static microwedge}

\begin{figure}[t]
\centering
\includegraphics[clip=,width= 1 \columnwidth]{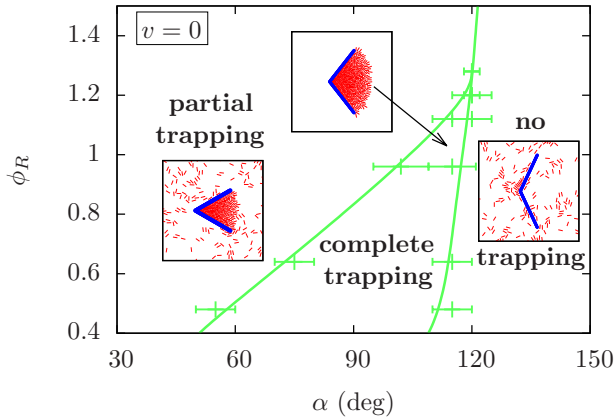}
\caption{(color online) Trapping state diagram in the case of a static trap $v=0$ marking three different states for varying
apex angle and reduced self-propelled rod packing fraction. All occurring trapping states are visualized 
by characteristic snapshots using central sections of the simulation box.
\label{f3}
}
\end{figure}

Let us first briefly recapitulate previous results \cite{Kaiser_PRL} for a static trap ($v=0$).
The trapping state diagram now drawn in the parameter space spanned by the net apex angle $\alpha$ 
and reduced rod packing fraction $\phi_{R}$ at fixed net densities is shown in Fig.~\ref{f3}, including characteristic snapshots. 

Following earlier work \cite{Kaiser_PRL}, we consider a rod $\alpha$ trapped, if its velocity $v_{\alpha} = | \textbf{v}_{\alpha} | < 0.1v_0$ for a time interval $t^{\ast} = 25\tau$. In contrast to the single particle case we now have to distinguish between two different kinds of trapped states. These are characterized
by the fraction of trapped particles $x_{T}$ which acts as some kind of order parameter for the different states.
Either no particle is trapped, $x_{T}=0$ (no trapping), or all particles are trapped, $x_{T}=1$ (complete trapping), 
or just a fraction of all particles in the system can be captured by the wedge, $0 < x_{T} < 1$. 
This new state will be referred to as partial trapping.

All trapping states can be found in the state diagram for a static microwedge, see Fig.~\ref{f3}. Evidently, 
in case of small apex angles there is only partial trapping 
since the area of the wedge is too small to accommodate all particles.

Larger apex angles enable complete trapping up to a certain reduced rod density. The resulting triple point
is independent of the trap density and attains a value $\phi_{R}^{\ast} \approx 1.3$. Higher densities will only show two different trapping states.
 While for a single particle and a static microwedge a capture is only possible for an apex angle $\alpha < 90^{\circ}$, an increase of the rod density leads to an increase of the maximum apex angle which allows trapping. Self-propelled rods coherently self-trap at the kink of the trap and jam.
Hereby a small immobile cluster can be formed which grows and leads to a filling of the wedge.
 Therefore, in the studied density regime, a trapping state is possible for apex angles up to $\alpha \approx 120^{\circ}$. The influence of rotational noise, which may arise from fluctuations in the swimming direction as manifested by {\em run-and-tumble} motion of swimming bacteria, has been accounted for by adding Gaussian white noise to the equation of rotational motion \equ{e:eom_u}. No significant effect on the trapping state diagram was found for typical values of the effective rotational diffusivity of bacterial swimmers \cite{Kaiser_PRL}.

\subsection{Mobile microwedge}

\begin{figure*}[t]
\centering
\includegraphics[clip=,width= 1.95 \columnwidth]{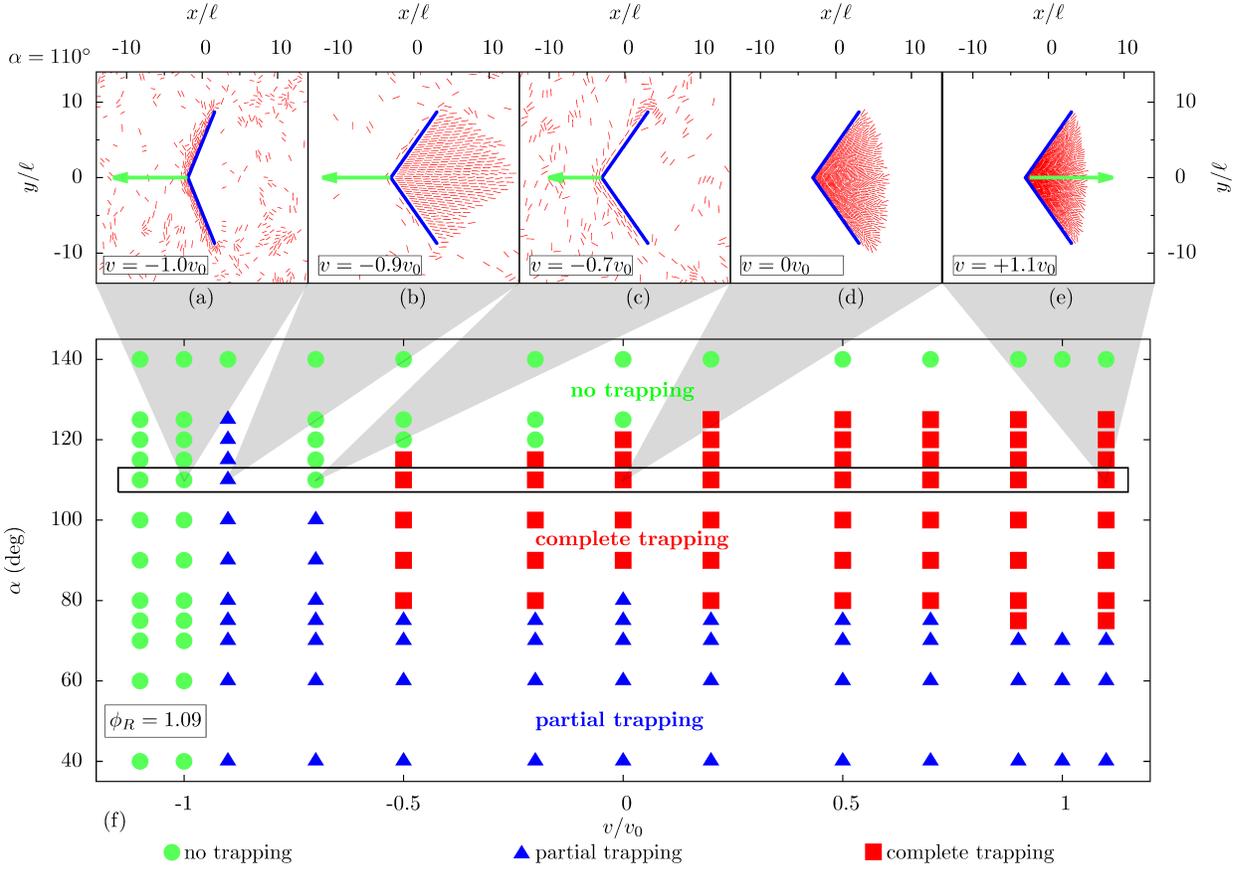}
\caption{(color online) Trapping state diagram and simulation snapshots of the final state at finite rod density $\phi_{R}=1.09$.
(a) - (e) Simulation snapshots for an apex angle $\alpha=110^{\circ}$. The respective dragging velocity of the trap is given
in each figure and indicated by a scaled arrow.
(f)  State diagram showing the three different trapping states in the plane spanned by the reduced net velocity and the trap
apex angle $\alpha$.
Circles correspond to no trapping, triangles to partial trapping and squares to complete trapping.
\label{f4}
}
\end{figure*}

We now consider a moving trap. Simulation results for the trapping state diagram in the
 plane spanned by reduced trap velocity $v/v_{0}$ and apex angle $\alpha$
are shown in Fig.~\ref{f4} combined with appropriate simulation snapshots characterizing the final state.
As a first general finding, the state diagram
has the same topology as that for a single rod if
one does not discriminate between partial trapping and complete trapping.
Of course, the actual numbers for the trapping to
no trapping state are significantly shifted. In particular, the worst case of trapping which occurs at an
opening angle of $90^{\circ}$ for a static microwedge, (see Fig.~\ref{f2}(a)), is now significantly shifted towards
an opening angle of about $110^{\circ}$ at a negative reduced drag velocity of about -0.7.
A corresponding snapshot of the empty microwedge is shown in Fig.~\ref{f4}(c).

If the trap velocity is varied at a fixed opening angle of $110^{\circ}$, as indicated by the different snapshots in Fig. \ref{f4}(a)-(e),
there is an intermediate trapping effect at negative drags close to -1 as indicated in Fig. \ref{f4}(b).
In this case rods can catch up with the moving net to accumulate inside the wedge. This is opposed to
strongly negative dragging velocities $v \leq - v_{0}$ where the wedge is faster than the rods on average
which leads to an accumulation of particles outside the net (see Fig.~\ref{f4}(a)).

Let us focus  on the partial trapping situation of rods which are only slightly faster than the net
as shown in Fig.~\ref{f4}(b). We observe a large swarm following the movement of net. The structure of the swarm is characterized by a significant
degree of nematic (or polar) order which stems from the repulsive rod interactions. The big swarm is therefore
a result of rod self-assembly templated by the moving net. The net plays the role of a leader which guides the swarm.
This is an interesting collective effect which can be in principle exploited to control and guide assemblies of  active particles
at wish \cite{Aguilar_Loewen_Yeomans_EPJE_2012} or to align them dynamically in an efficient way. Qualitatively, the tendency
of aligment can be read off already from a single rod (see Fig.~\ref{f2}(c)) which tries to orient along the drag direction.
The rod interaction, however, dramatically increases the alignment leading to a large orientated swarm.

Further increasing positive drag speed will compress the trapped particles leading to a larger number density
of the captured particles inside the net. Therefore at higher speeds the threshold to a no trapping state
is shifted towards larger opening angles.

We now focus on the transition line between partial trapping and complete trapping, 
see the squares and triangles in Fig.~\ref{f4}(f). At fixed opening angle
(say at about $80^{\circ}$), this line
also shows an interesting reentrance effect for increasing drag velocities as embodied in the highly
nontrivial state sequence
 partial trapping -  complete trapping -  partial trapping -  complete trapping.
The first transition from  partial trapping to  complete trapping has to do with the efficient nematization
which then gets less efficient at almost zero drag velocities. The ultimative transition to complete trapping
is then an effect of rod compression inside the net for increasing drag velocities.
Interestingly, starting with a resting net with opening angles slightly below $90^\circ$ degrees, the trapping efficiency
increases no matter in which direction the microwedge is dragged.

\begin{figure}[t]
\centering
\includegraphics[clip=,width= 1 \columnwidth]{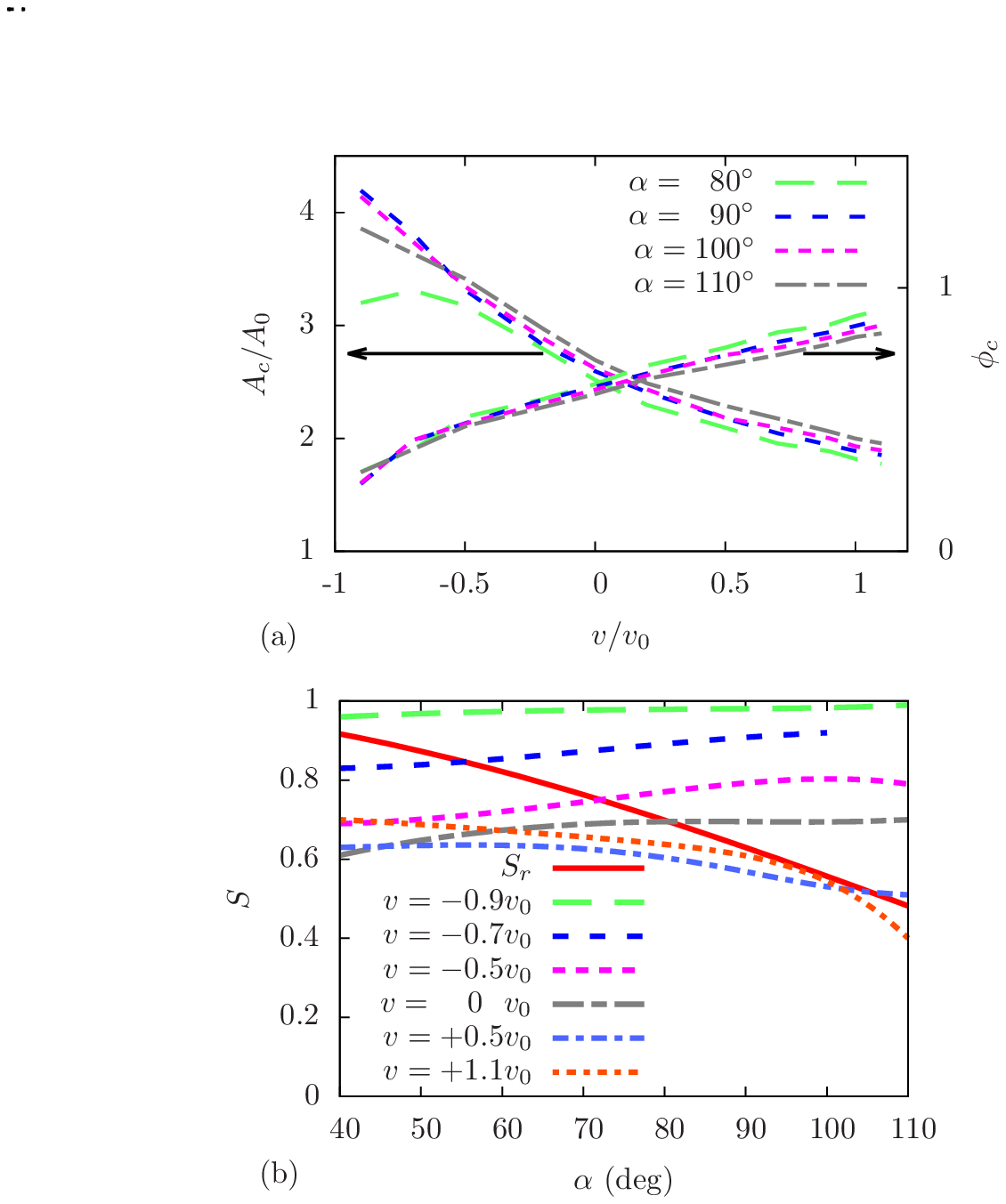}
\caption{(color online) (a) Relative area occupied by captured swimmers $A_c / A_0$ and resulting packing fraction $\phi_c$  for $\phi_{R} = 1.09$ and varying drag velocities for four apex angles.
(b) Dependence of the nematic order parameter $S$ on the apex angle for various drag velocities (dashed lines). The reference value  $S_{r}$  for a perfect cone-orientation of the captured rods is given by the solid line.
\label{f5}
}
\end{figure}

\begin{figure}[htp]
\begin{center}
\includegraphics[width=1\columnwidth]{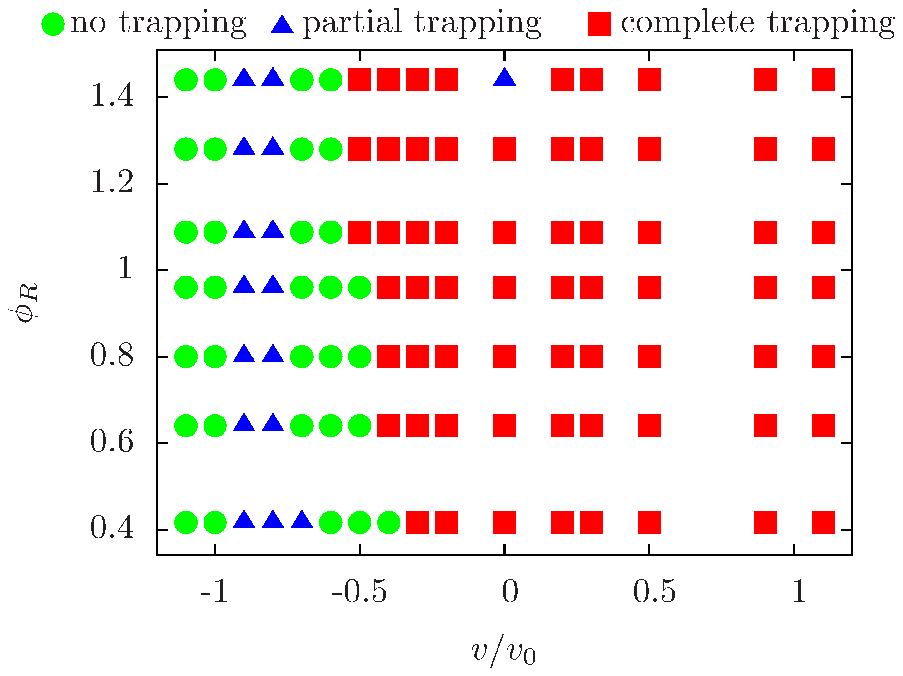}
\end{center}
\caption{(color online) Trapping states for a fixed apex angle $\alpha = 110^{\circ}$.
\label{f6}
}
\end{figure}

We now characterize the directed self-assembled state more carefully by monitoring
the area
covered by the trapped particles and the actual nematic order. First we
draw a convex hull
around all trapped particles which defines an area $A_c$. We normalize
this area to the inside area
$A_0= (\ell_T /8) \sin \alpha/2$ of the wedge. Results for $A_c/A_0$ as a
function of the
drag speed are presented in Fig.~\ref{f5}(a) at fixed opening angle $\alpha$.
In line with the huge nematic
wake discussed earlier, the ratio $A_c/A_0$ vastly exceeds unity for
negative drags  close to $-v_0$. In fact,
$A_c/A_0$ has a maximum as a function of $v/v_0$ which points to a very
efficient wake area that contains
particles which are dragged through the liquid by the moving wedge.

Second, we analyze the degree of nematic ordering in the trapped particles
by calculating the average
\bea
S = \langle 2 \cos^{2} \theta_{i}  -1 \rangle,
\eea
where $\theta_i$ is the angle between the rod orientation of the $i$th rod
and the drag velocity.
The average $\langle \ldots \rangle $ refers to an average over all captured rods for a
variety of different initial
configurations. The nematic order parameter $S$ is defined as usual in two
spatial dimensions.
For a perfect alignment of all trapped rods, $S=1$, while $S$ vanishes if
there is no
orientational ordering at all. We relate this quantity $S$ to a perfect
cone filling of the rods
where the orientational direction is anti-radially towards the origin of
the wedge. In this reference
situation, the nematic order parameter $S_r$ is given by
\bea
S_{r} \ = \frac{1}{\alpha} \int_{-\alpha /2}^{\alpha /2}  (2 \cos ^{2} \theta -1 ) d \theta = \sin \alpha/ \alpha.
\eea

In Fig.~\ref{f5}(b), $S$ is shown versus the opening angle for fixed drag speeds.
The cone normalization $S_r$ is also given. For the nematic swarm at
$v/v_0 =-0.9$, $S$ clearly
exceeds $S_r$. This is inverted for very high positive drags $v>v_0$,
where $S<S_r$ holds
over the full range of opening angles. This finding can be attributed
mainly to particle
misorientations at the wedge boundary close to the end of the wedge, see
again the
snapshot of Fig.~\ref{f4}(e).

In addition, we consider a system with a fixed apex angle $\alpha = 110^{\circ}$ and vary the reduced rod packing fraction
$\phi_{R}$ and the dragging velocity. The data contained in Fig.~\ref{f6} show that the dependence on the rod density is weak
providing the same state-sequence as for the special rod density selected previously for Fig.~\ref{f4}(f).
Only, in the case of extreme rod densities $\phi_R > \phi_{R}^{\ast}=1.3$, the area of the net is not large enough to accomodate all particles.
This leads to a partial trapping state instead of complete trapping as indicated in Fig.~\ref{f6} for a static microwedge
at high rod densities.

We expect our results to be robust against hydrodynamic far-field interactions which are expected to be less important for the particle dynamics at high local particle densities, as  found inside the trap, due to mutual hydrodynamic screening \cite{Muthu} and the small magnitude of the  flows fields generated by the  microswimmers \cite{DrescherPNAS} and the moving wedge. Moreover, the presence of no-slip trap boundaries in microfluidic devices are expected to strongly suppress hydrodynamic long-range interactions between swimmers due to cancellation effects from the hydrodynamic images \cite{Wensink_PNAS}.

\section{Conclusions}
\label{sec:conc}
While there is a considerable amount of detailed knowledge about how to trap macroscopic particles and passive particles in static traps such as
colloids using optical tweezers or atoms in a Paul trap, is it much less clear how large numbers of active microscopic particles can be captured in an efficient way.
Using computer simulations we have studied a dragged chevron-shaped trap which allows to capture several self-propelled rods in an irreversible manner.
A microwedge with variable apex angle $\alpha$ enforces active particles to rectify their swimming
direction and stimulates the formation of microscopic cluster which may subsequently act as a nucleus for a fast-growing mesoscopic aggregate of captured rods. We have demonstrated the crucial role of the apex angle and the drag velocity of the trap. A non-zero drag velocity imposes dynamic nematisation and layer-like ordering of the clustered rods provided the drag velocity is slightly above $-v_0$ ($ v \gtrsim -v_0$).
We have highlighted the influence of collective self-trapping by comparing results for many self-propelled rods with the single particle case.
The dramatical collective response of self-propelled rods to a minor change in the boundary shape or drag velocity is 
remarkable and remains unseen for passive systems exposed to external boundaries or electromagnetic traps.

Collective trapping of ensembles of active particles in moving traps can be verified by experiments using rod-shaped bacteria \cite{cisnerosgoldstein}
 or driven polar granular rods \cite{kudrolli} exposed to geometrically structured boundaries \cite{Bechinger_SM11,microfluidic,Mino_Clement}.
While the presented results are valid for linearly propagating swimmers it would be interesting to
study a trapping device for several swimmers moving on circle-like pathways \cite{BtH_L_part,Kaiser_PRE,SvT2009,VolpeARXIV}.
Furthermore it would be interesting to exploit the trapping scenarios proposed here to design a trapping device which is capable of extracting swimmers with a specific velocity bandwidth from a mixture of active particles with a strong spread in motility. According to the results in Fig.~\ref{f4} such a velocity selective trapping could be envisaged by dragging the net at a judiciously chosen negative drag velocity such as to facilitate templated clustering of a subset of swimmers whose individual motility closely resembles that of the moving net.

An interesting open question is to which extent the details of the propulsion mechanism have 
an impact on the self-trapping behaviour of rods.
In particular, it would be interesting to study whether puller and pusher-type swimmers exhibit different trapping behaviour.
These problems will necessitate the use of more sophisticated simulation schemes \cite{Ishikawa2008,squirmerStark,gompper} and
 brings us also to the question about the importance of hydrodynamic near-field interactions 
\cite{Dhontbook} which are ignored in our model. Real bacteria are usually propelled by flagella attached to the bacterial body whose internal configuration will presumably change at high bacterial density, under strong confinement or at an obstacle {\cite{Goldstein_PRE2006R}.
These flagellar interactions may lead to more specific effects which are neglected in our model but could be included on a coarse-grained level  in future studies. In particular, one could introduce a density-dependent microscopic mobility which is known to have a considerable effect on the collective behaviour  in bulk \cite{Marchetti_PRL2012}.
Finally it would be interesting to model the properties of the trapped polar state of rods using continuum  elasticity theory
following recent efforts in this direction for the wetting behaviour of (passive) liquid crystals
 confined in wedges \cite{LettingaPRL2012,daGama,Parry}. 

\acknowledgments 
We thank J. Tailleur  for helpful discussions.
This work was financially supported by the ERC Advanced Grant INTERCOCOS
(grant agreement 267499) and by the DFG within SFB TR6 (project D3).

\bibliographystyle{apsrev}

\bibliography{refs}

\end{document}